\documentclass[aps,pre,twocolumn,groupedaddress,superscriptaddress,showpacs]{revtex4-1}

\usepackage{graphicx}
\usepackage{color}
\usepackage{amsmath}
\usepackage{amsfonts}
\usepackage{amssymb}
\usepackage{dcolumn}
\usepackage{hyperref}
\usepackage{enumerate}
\usepackage{bbold}
\usepackage{cancel}
\usepackage{mathtools}
\usepackage[]{algorithm2e}
\usepackage{siunitx}
\usepackage{cleveref}
\newcommand*\widebar[1]{%
   \hbox{%
     \vbox{%
       \hrule height 0.5pt 
       \kern0.5ex
       \hbox{%
         \kern-0.1em
         \ensuremath{#1}%
         \kern-0.1em
       }%
     }%
   }
} 
\DeclarePairedDelimiter\floor{\lfloor}{\rfloor}

\usepackage{overpic}

\begin{document}

\title{Unifying continuous, discrete, and hybrid susceptible-infected-recovered processes on networks}
\date{\today}
\author{Lucas B\"ottcher}
\affiliation{Computational Medicine, UCLA, 90095-1766, Los Angeles, United States}
\affiliation{Institute for Theoretical Physics, ETH Zurich, 8093, Zurich, Switzerland}
\affiliation{Center of Economic Research, ETH Zurich, 8092, Zurich, Switzerland}
\email{lucasb@ethz.ch}
\author{Nino Antulov-Fantulin}
\affiliation{Computational Social Science, ETH Zurich, 8092, Zurich, Switzerland}
\email{anino@ethz.ch}
\date{\today}
\begin{abstract}
Waiting times between two consecutive infection and recovery events in spreading processes are often assumed to be exponentially distributed, which results in Markovian (i.e., memoryless) continuous spreading dynamics. However, this is not taking into account memory (correlation) effects and discrete interactions that have been identified as relevant in social, transportation, and disease dynamics. We introduce a framework to model continuous, discrete, and hybrid forms of (non-)Markovian susceptible-infected-recovered (SIR) stochastic processes on networks. The hybrid SIR processes that we study in this paper describe infections as discrete-time Markovian and recovery events as continuous-time non-Markovian processes, which mimic the distribution of cell cycles. Our results suggest that the effective-infection-rate description of epidemic processes fails to uniquely capture the behavior of such hybrid and also general non-Markovian disease dynamics. Providing a unifying description of general Markovian and non-Markovian disease outbreaks, we instead show that the mean transmissibility produces the same phase diagrams independent of the underlying inter-event-time distributions.
\end{abstract}
\maketitle
\section{Introduction}
Models of epidemic processes such as the \emph{susceptible-infected-recovered} (SIR) model and related models provided various insights into dynamical and stationary features of disease, opinion, and failure spread in social and technical systems~\cite{pastor2015epidemic,bottcher2017critical, SIR_PRL_InverseProblem}. In the SIR model, infected individuals may transmit a disease to susceptible ones. After a certain period, infected individuals recover and are not part of the disease-transmission process anymore. 
The exact time evolution of the continuous-time stochastic SIR spreading process is described by the Chapman-Kolmogorov equation or its differential form (i.e., the master equation). However, exact analytical solutions of the master equation are limited to special cases and therefore different approximations are being used (e.g., deterministic ODE models~\cite{moreno2002epidemic, keeling2011modeling, sharkey2011deterministic}, cavity-like models~\cite{sharkey2015exact, karrer2010message}, and pairwise approaches~\cite{sherborne2018mean}). 

Gillespie and \emph{kinetic Monte-Carlo} (kMC) approaches~\cite{gillespie1976general} provide techniques to generate statistically exact trajectories of a master equation. The assumption underlying kMC simulations of SIR processes is that waiting times between consecutive recovery and infection events are exponentially distributed. However, many natural processes including social dynamics~\cite{barabasi2005origin,bottcher2017temporal} exhibit correlation and memory (i.e., non-Markovian) effects~\cite{goh2008burstiness} and are therefore not described by exponential (i.e., memoryless) waiting-time distributions~\cite{starnini2017equivalence}. Recent attempts to simulate general non-Markovian processes led to the development of the \emph{non-Markovian Gillespie algorithm} (nMGA)~\cite{boguna2014simulating} and the \emph{Laplace Gillespie algorithm} (LGA)~\cite{masuda2018gillespie}. Both methods are based on a mapping of multiple stochastic processes with general (continuous) waiting-time distributions to a modified kMC algorithm. As outlined in Ref.~\cite{masuda2018gillespie}, the nMGA is exact only for infinitely many processes and requires to re-calculate all individual rates at every time step. This algorithm has the advantage that it can simulate arbitrary continuous inter-event-time distributions~\cite{boguna2014simulating}. The LGA interprets survival functions of waiting-time distributions as Laplace transforms of underlying event-rate distributions. Although the LGA is exact for arbitrary numbers of processes, it is only applicable to certain waiting-time distributions~\cite{masuda2018gillespie}.

In the context of non-Markovian SIR models, there also exist event-driven and directed-percolation-based approaches that are directly applicable to these types of processes~\cite{tolic2018simulating,istvan2019mathematics}. In this paper, we use an approach similar to the directed-percolation method of Refs.~\cite{tolic2018simulating,istvan2019mathematics} and map non-Markovian SIR processes with general waiting-time distributions to a shortest-path problem~\cite{kulkarni1986shortest,corea1993shortest} in a weighted spreading network. We refer to this method as \emph{shortest-path kinetic Monte Carlo} (SPkMC). In contrast to the nMGA, LGA, and aforementioned approximation techniques~\cite{moreno2002epidemic, keeling2011modeling, sharkey2011deterministic,sharkey2015exact, karrer2010message,sherborne2018mean}, our SPkMC framework allows us to produce exact stochastic trajectories of SIR processes for general continuous and discrete waiting-time distributions on any network.
\begin{figure*}
\centering
\includegraphics[width=0.75\textwidth]{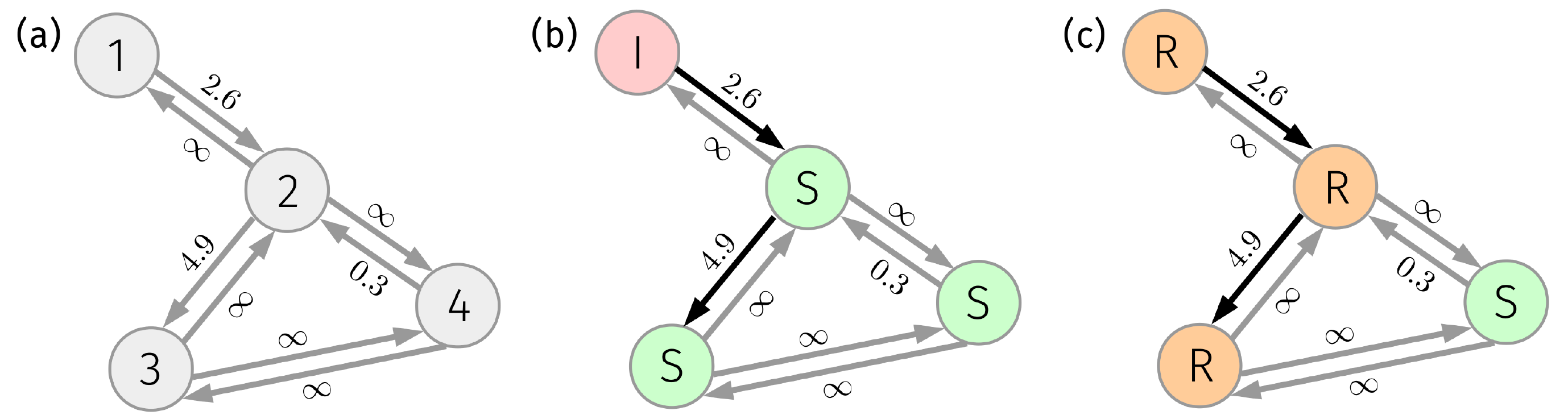}
\caption{\textbf{SIR dynamics on a spreading network.} (a) Initialization of edge weights according to Eq.~\eqref{eq:weights} in a spreading network that consists of 4 nodes. Finite edge weights indicate that disease transmission can occur along the corresponding edge. (b) Node 1 is infected and transmits the disease to susceptible nodes 2 and 3 that are connected via paths of finite length (indicated by black arrows). The times at which nodes 2 and 3 get infected are $\rho_{12}=2.6$ and $\rho_{12}+\rho_{23}=7.5$. All paths that connect node 1 with node 4 have infinite length, so node 4 cannot be infected. (c) In the long-time limit $\tau \rightarrow \infty$, all infected nodes recover.}
\label{fig:spreading_network}
\end{figure*}

We use our framework to study continuous, discrete, and hybrid SIR processes. Such hybrid formulations of spreading dynamics can be useful to account for events that are generated according to a sequence of fixed-duration schedules (e.g., meeting~\cite{panko1995meeting}, patient care~\cite{rea2007duration,duval2018measuring}, and transportation schedules~\cite{dorfman2004scheduling}) and (ii) finite transfer times of information in communication networks~\cite{chen2003modeling}.

Various processes that affect human interaction and information exchange in communication networks can be modeled as discrete events. Possible examples include sequences of fixed-duration schedules of meetings (multiples of hours and minutes)~\cite{panko1995meeting,rea2007duration,duval2018measuring} and transportation schedules~\cite{dorfman2004scheduling} that can affect disease transmission on a meta-population level. For communication networks, finite transfer times and synced digital events (e.g., release of computer viruses) are also examples of discrete processes~\cite{chen2003modeling}.

In addition, hybrid SIR models can also account for latency periods that are typically modeled by introducing an additional ``exposed'' compartment~\cite{keeling2011modeling} and a corresponding latent period, which mimics the observed mean incubation time (e.g., 8-14 days for measles~\cite{lessler2009incubation}). As an alternative to discrete~\cite{allen1994some,zhou2004discrete,chen2019discrete} and continuous~\cite{keeling2011modeling} epidemic models with ``exposed'' compartments, our approach can account for discrete time delays by directly modifying transmission rates.

Our results suggest that hybrid and also general non-Markovian disease outbreaks cannot be uniquely captured by effective infection rates. However, by mapping hybrid SIR processes to bond percolation~\cite{grassberger1983critical,newman2002spread}, we show that the corresponding mean transmissibilities produce the same phase diagrams independent of the underlying infection- and recovery-time distributions; thus providing a unifying description of general Markovian and non-Markovian SIR processes.
The framework we propose assumes no specific form of waiting-time distributions~\cite{chen2019discrete,fennell2016limitations} and allows us to simulate and analytically describe general discrete, continuous, and hybrid variants of Markovian and non-Markovian SIR dynamics.
\section{Shortest-path kinetic Monte Carlo}
\label{sec:spkmc}
Before focusing on the simulation of hybrid SIR dynamics, we introduce the necessary mathematical toolbox that allows us to map general waiting times to a shortest-path problem in an underlying spreading network.
Let $\phi(\tau)$ and $\psi(\tau)$ be the probability-density or probability mass functions (PDFs or PMFs) of recovery and infection times. In continuous time, the probability that a recovery (infection) event occurs in the interval $(\tau,\tau+\mathrm{d} \tau)$ is $\phi(\tau)\, \mathrm{d}\tau$ ($\psi(\tau)\, \mathrm{d}\tau$). The discrete time analogues $\phi(\tau)$ and $\psi(\tau)$ are the recovery and infection probabilities after $\floor{\tau}$ steps, where $\floor{\cdot}$ denotes the floor function. We denote the cumulative distribution function (CDF) of $\phi(\tau)$ and $\psi(\tau)$ by $\Phi(\tau)$ and $\Psi(\tau)$. The function $\Phi(\tau)$ ($\Psi(\tau)$) is the probability that a recovery (infection) event occurred in $[0,\tau]$.  In the case of Poissonian SIR dynamics, waiting-time distributions are described by the PDFs $\phi(\tau)=\gamma e^{-\gamma \tau}$ and $\psi(\tau)=\beta e^{-\beta \tau}$ and CDFs $\Phi(\tau)=1-e^{-\gamma \tau}$ and $\Psi(\tau)=1-e^{-\beta \tau}$, where $\gamma$ and $\beta$ are the corresponding recovery and infection rates. Note  that we use prefixes such as ``Erlang-geometric'' to indicate the recovery and infection time distributions ($\phi(\tau)$ and $\psi(\tau)$) of the corresponding hybrid SIR process. 

For given distributions $\phi(\tau)$ and $\psi(\tau)$, we consider $M$ realizations of SIR dynamics to correspond to an ensemble of $M$ directed spreading networks $\{G_k(V,E)\}_{k\in\{1,\dots,M\}}$, where $V$ and $E$ denote the sets of nodes and edges.
Each network $G_k(V,E)$ is initialized as follows. For each node $s$ in $G_k(V,E)$, we generate a random number $x\sim \mathcal{U}(0,1)$ and use an inverse transform sampling of $\Phi(\tau)$ to determine the recovery time of node $s$ according to $\Phi^{-1}(x)$. For each node $t$ that is adjacent to $s$, we generate another random number $y \sim \mathcal{U}(0,1)$ and determine the infection time $\Psi^{-1}(y)$. We now use $\Phi^{-1}(x)$ and $\Psi^{-1}(y)$ to determine edge weights~\cite{tolic2018simulating}
\begin{equation}
\rho_{s t} = 
\begin{cases}
\Psi^{-1}(y) \,, &\Psi^{-1}(y) \leq \Phi^{-1}(x)\,, \\
\infty \,, &\Psi^{-1}(y) > \Phi^{-1}(x) \,.
\end{cases}
\label{eq:weights}
\end{equation}
We set $\rho_{s t}$ to $\Psi^{-1}(y)$ (i.e., the disease transmission time from node $s$ to $t$) if infection occurs before recovery, and $\rho_{s t}=\infty$ otherwise. Note that the interaction terms can also be general $\rho_{s t} = f(\Psi(y), \Phi(x), \theta_s, \theta_t)$, where $f(\cdot)$ accounts for node-dependent transmission features $(\theta_s, \theta_t)$ like age, gender, and other social and demographic factors including interventions like the probability of quarantine or contact containment restrictions. If the CDFs are not invertible, we can generate edge weights with rejection sampling. In Fig.~\ref{fig:spreading_network} (a), we show an illustration of the weight initialization procedure for a network that consists of 4 nodes. We again note that networks $G_k(V,E)$ are directed (i.e., weights $\rho_{st}$ may be different from $\rho_{ts}$). In the case of Poissonian dynamics, we obtain
\begin{equation}
\rho_{s t} = 
\begin{cases}
-\frac{\ln({x})}{\beta} \,, &-\frac{\ln({x})}{\beta}   \leq -\frac{\ln({y})}{\gamma}  \,, \\
\infty \,, &-\frac{\ln({x})}{\beta}   > -\frac{\ln({y})}{\gamma}\,.
\end{cases}
\label{eq:Poissonian}
\end{equation}
In addition to edge weights, we also keep track of node weights $\tau_i=\Phi^{-1}(x)$ to describe the evolution of SIR dynamics in a network. 

\begin{figure}
\centering
\includegraphics[width=0.42\textwidth]{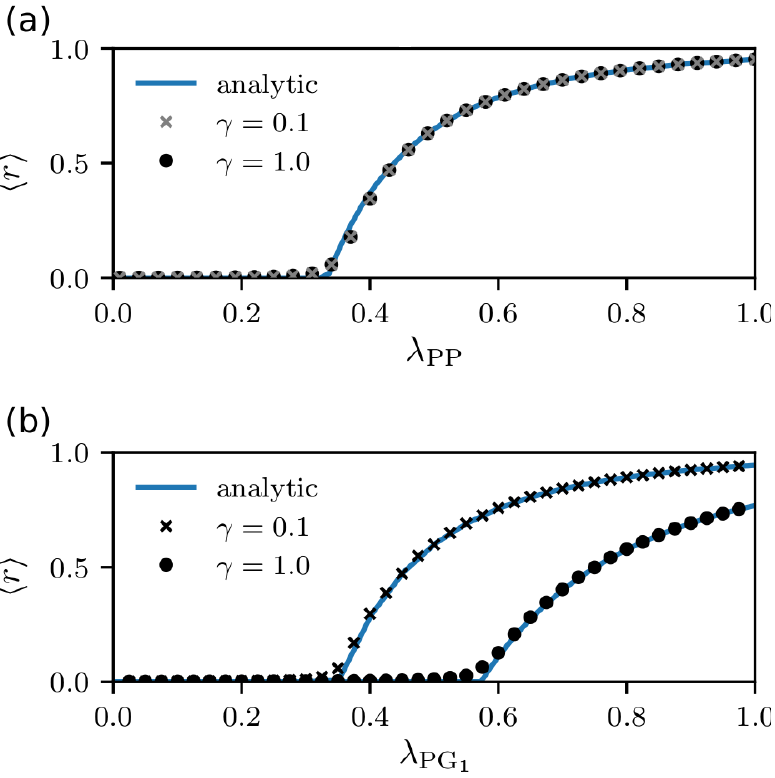}
\caption{\textbf{Fraction of recovered nodes in Poissonian and 
hybrid Poisson-geometric SIR processes.}  The fraction of recovered nodes $\langle r \rangle$ (see Eq.~\eqref{eq:r_ensemble}) as a function of the effective infection rate for (a) fully Poissonian and (b) Poisson-geometric) SIR processes. The effective spreading rates are $\lambda_{\mathrm{PP}}=\beta/\gamma$ and $\lambda_{\mathrm{PG}_1}=q/\gamma$. Analytical solutions (blue solid lines) are based on Eqs.~\eqref{eq:outbreak_size} and \eqref{eq:generating_functions} for a random-regular graph with $k=5$. Numerical simulations have been performed for $N=10^5$ nodes and $M=10^2$ realizations.}
\label{fig:hybrid_outbreak}
\end{figure}

After having identified all weights, we obtain one realization $G_k(V,E)$ of the spreading network. In the next step, we take $G_k(V,E)$ and infect one uniformly at random selected node (see Fig.~\ref{fig:spreading_network} (b)). All nodes that are connected to the initially infected node through paths of finite length are also infected and recover in the limit of $\tau \rightarrow \infty$ (see Fig.~\ref{fig:spreading_network} (c)). The shortest-path length between an infected source and its target node is the corresponding disease transmission time. This formulation of disease transmission can be viewed as a ``least action principle'' for kMC.
If all paths that connect two nodes are infinite, we know that one or multiple recovered (or quarantined/removed) nodes hinder disease transmission (see node 4 in Fig.~\ref{fig:spreading_network}). To describe SIR dynamics with $n$ initially infected nodes, we use $I_k^j$ to denote the set of infected nodes that result from an initial infection of node $j$ in $G_k(E,V)$. The corresponding set of all infected nodes that result from multiple spreading seeds in $G_k(V,E)$ is $I_k = \bigcup_{j=1}^n I_k^j$. The SPkMC framework also allows us to monitor the infection and recovery times of individual nodes. In App.~\ref{app:dynamics} and App.~\ref{app:QuarantineDynamics}, we outline how the evolution of susceptible, infected, and recovered nodes can be reconstructed from shortest paths and describe the possibility to account for quarantine protocols in SPkMC simulations.

The stationary fraction of recovered nodes in network $G_k(E,V)$ is $r_k=|I_k|/N$, where $N=|V|$ is the number of nodes, and the corresponding ensemble average over $\{G_k(V,E)\}_{k\in\{1,\dots,M\}}$ yields
\begin{equation}
\langle r \rangle=\frac{1}{M} \sum_{k=1}^M r_k\,.
\label{eq:r_ensemble}
\end{equation}
For each network realization $G_k(E,V)$, we have to identify all shortest paths, an operation with run time of order $\mathcal{O}(|E|+|V| \log |V|)$ when using optimized data structures such as Fibonacci heaps~\cite{fredman1987fibonacci}. For $n$ initially infected nodes, we can determine the number of susceptible $S(t)$, infected $I(t)$, and recovered $R(t)$ nodes at time $t$ (see App.~\ref{app:dynamics}) after running Dijkstra's algorithm $n$ times. Typically, the number of initially infected nodes $n$ is small and thus the run time of our algorithm is still of order $\mathcal{O}(|E|+|V| \log |V|)$. Reference \cite{masuda2018gillespie} discusses the run time complexity of the nMGA and LGA \emph{per generated event}. Since our SPkMC framework can simulate SIR dynamics with one run of Dijkstra's algorithm, the computational complexity of our framework does not depend on the number of time steps that one wants to simulate.

In App.~\ref{app:comparison}, we consider Poissonian dynamics (see Eq.~\eqref{eq:Poissonian}) and show that our SPkMC simulations of stationary and dynamical SIR features agree well with corresponding kMC simulations. An advantage of our proposed shortest-path SIR simulation method is the possibility to simulate general Markovian and non-Markovian dynamics with continuous and discrete waiting time distributions.
\section{Hybrid continuous-discrete SIR dynamics}
\label{sec:hybrid}
To describe latency periods in infection processes (i.e., no infection occurs during a certain time window), we apply our simulation framework to hybrid Poisson-geometric SIR dynamics with discrete infection events that are distributed according to a geometric density function
\begin{equation}
\psi_{\mathrm{G}_1}(\tau)=\sum_{k=1}^{\infty} \delta(\tau-k) (1-q)^{k-1}q\,,
\label{eq:geometric}
\end{equation}
where $q$ is the probability that an infection event occurs within a time step of 1 and $\delta(\cdot)$ is the Dirac delta function. We show in App.~\ref{sec:master_equation} that the master equation of hybrid Poisson-geometric SIR dynamics is no longer time-homogeneous. In App.~\ref{app:discrete_inf_times}, we also consider an alternative definition of the geometric distribution $\psi_{\mathrm{G}_2}$ that takes on finite values for all non-negative integers. The geometric distribution is the discrete memoryless counterpart of exponential distributions.

We can now use our simulation framework to study disease outbreak characteristics of such hybrid SIR processes. As for many epidemic processes~\cite{keeling2011modeling}, we characterize disease dynamics in terms of the effective infection rate $\lambda=\langle \tau \rangle_{\phi}/\langle \tau \rangle_{\psi}$, where $\langle \tau \rangle_{\psi}$ and $\langle \tau \rangle_{\phi}$ are the mean times to infection and recovery, respectively. For a fully Poissonian SIR process, the effective infection rate is $\lambda_{\mathrm{PP}}=\beta/\gamma$ and invariant upon rescaling of infection and recovery rates by a constant factor~\cite{pastor2015epidemic}. That is, the corresponding fraction of recovered (see Eq.~\eqref{eq:r_ensemble}) only depends on the effective infection rate $\lambda_{\mathrm{PP}}$ (see Fig.~\ref{fig:hybrid_outbreak} (top)). By analogy, we now use $\lambda_{\mathrm{PG}_1}=1/(\gamma \langle \tau \rangle_{\psi})$ to denote the effective infection rate for Poissonian-geometric dynamics with infection-time PDF $\psi_{\mathrm{G}_1}$ and
\begin{equation}
\langle \tau \rangle_{\psi} = \sum_{\tau'=1}^\infty \tau' (1-q)^{\tau'-1} q=q^{-1}\,.
\label{eq:mean_tau}
\end{equation}
However, unlike in fully Poissonian SIR dynamics, we cannot uniquely capture the corresponding phase space by the effective infection rate $\lambda_{\mathrm{PG}_1}$ (see Fig.~\ref{fig:hybrid_outbreak} (bottom)). That is, we observe different fractions of recovered $\langle r \rangle$ for the same value of $\lambda_{\mathrm{PG}_1}$. To better understand the phase space of hybrid SIR processes, we proceed with a mapping to bond percolation.
\section{Mapping hybrid SIR dynamics to bond percolation}
\label{sec:bond_perc}
We now analytically characterize the hybrid SIR disease prevalence in terms of the mean transmissibility $\widebar{T}$ that describes the probability of an infection to be transmitted from an infected to an adjacent susceptible node~\cite{newman2002spread}:
\begin{equation}
\widebar{T} = \int_0^\infty \phi(\tau) \int_0^\tau \psi(\tau^\prime)\, \mathrm{d}\tau^\prime\,\mathrm{d}\tau\,.
\label{eq:transmissibility}
\end{equation}
In networks with no degree correlations, the critical transmissibility above which an SIR epidemic spreads through a finite fraction of the system is given by the bond percolation threshold~\cite{newman2002spread,pastor2015epidemic}
\begin{equation}
p_c = \frac{\langle k \rangle}{\langle k^2 \rangle - \langle k \rangle}\,,
\label{eq:p_c}
\end{equation}
where $\langle k \rangle$ and $\langle k^2 \rangle$ denote the first and second moment of the degree distribution $P_k$.
\begin{figure}
\centering
\includegraphics[width=0.42\textwidth]{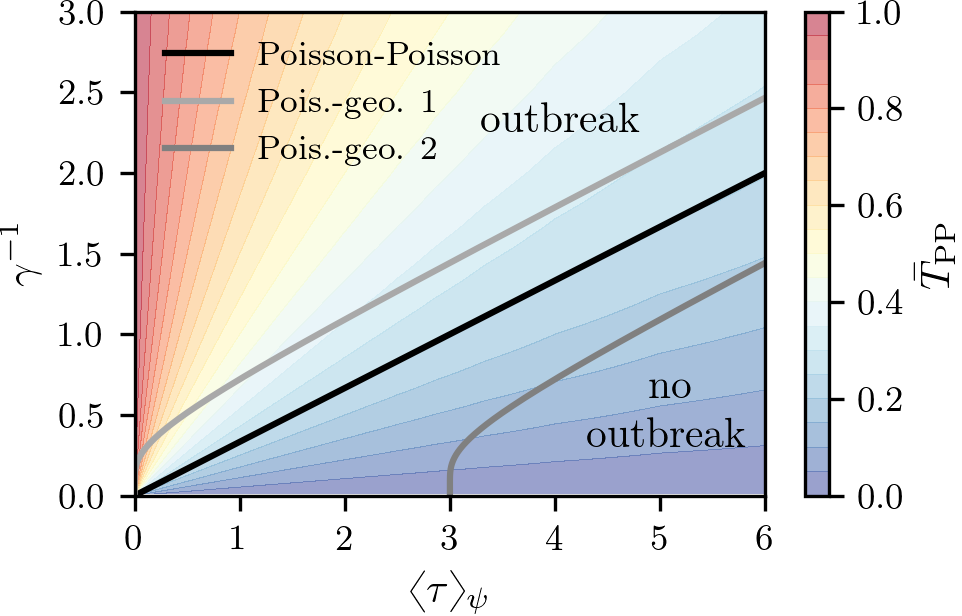}
\caption{\textbf{Comparison of phase spaces.} We show the separation lines between phases with and without disease outbreaks. The black solid line corresponds to Poissonian SIR dynamics with $T_{\mathrm{PP}}=\lambda_{\mathrm{PP}}/(1+\lambda_\mathrm{PP})$ and $\lambda_\mathrm{PP}=\beta/\gamma$. The light (dark) grey solid line describes the phase separation for Poissonian-geometric SIR dynamics with $\psi_{\mathrm{G}_1}$ ($\psi_{\mathrm{G}_2}$) (see Eqs.~\eqref{eq:geometric} and \eqref{eq:trans_g} and the SI for details).}
\label{fig:phase_space}
\end{figure}
For Poisson-geometric SIR dynamics, the mean transmissibility is
\begin{align}
\begin{split}
\widebar{T}_{\mathrm{PG}_1}&=\widebar{T}_{\mathrm{PG}_1}(\gamma, q) = e^{-\gamma}\left[1+\frac{\left(e^{\gamma }-1\right)  (q-1)}{e^{\gamma }+q-1}\right]\,.
\end{split}
\label{eq:trans_g}
\end{align}
In App.~\ref{app:discrete_inf_times}, we provide further details about the derivation of Eq.~\eqref{eq:trans_g} and compare the fully Poissonian and Poisson-geometric case. According to Eqs.~\eqref{eq:p_c} and \eqref{eq:trans_g}, we find the phase separation line $q_c=p_c(e^{\gamma}-1)/(1-p_c)$, which separates the phases with and without disease outbreaks (see Fig.~\ref{fig:phase_space}). For $\gamma=1$ and $\gamma=0.1$, we obtain the critical effective infection rates $\lambda_{\mathrm{PG}_1}^c\approx 0.57$ and $\lambda_{\mathrm{PG}_1}^c\approx 0.35$, respectively. These values agree well with the numerical data of Fig.~\ref{fig:hybrid_outbreak} (bottom).

Note that $\widebar{T}_{\mathrm{PG}_1}(\gamma,q)$ cannot be parametrized in terms of an effective infection rate $\lambda_{\mathrm{PG}_1}$ (see Fig.~\ref{fig:phase_space}) whereas for fully Poissonian dynamics the mean transmissibility $T_{\mathrm{PP}}=\lambda_{\mathrm{PP}}/(1+\lambda_\mathrm{PP})$ only depends on the effective infection rate $\lambda_\mathrm{PP}=\beta/\gamma$ (see App.~\ref{app:comparison} and Ref.~\cite{newman2002spread}).
However, if $\gamma$ is small, we find that 
\begin{equation}
\lim_{\gamma \rightarrow 0} \, \widebar{T}_{\mathrm{PG}_1}(\gamma, q)= \widebar{T}_{\mathrm{PG}_1}(\lambda_{\mathrm{PG}_1})= 1-\lambda_{\mathrm{PG}_1}^{-1}\,, 
\end{equation}
Thus, for sufficiently small values of $q$ and $\gamma$ (i.e., long mean infection and recovery times), the mean transmissibility $\widebar{T}_{\mathrm{PG}_1}(\gamma,q)$ only depends on the effective infection rate $\lambda_{\mathrm{PG}_1}$. In the SI, we show that this is also the case for the alternative formulation of Poisson-geometric SIR dynamics with $\psi_{\mathrm{G}_2}$. A graphical interpretation of this result is that the phase separation lines in Fig.~\ref{fig:phase_space} merge as $\langle \tau \rangle_\phi=\gamma^{-1}$ and $\langle \tau \rangle_\psi$ tend to infinity.

To determine the relative size of the epidemic $\mathcal{S}(\widebar{T})$ as function of the mean transmissibility $\widebar{T}$, we use a generating-function approach~\cite{newman2002spread} and first consider an uncorrelated network for which the conditional probability $P(k|k')= k P_k /\langle k \rangle$ does not depend on $k'$. This approach is based on two generating functions $G_0(x;\widebar{T})$ and $G_1(x;\widebar{T})$. The former is the generating function of the distribution of occupied edges belonging to a certain node, as a function of $\widebar{T}$~\cite{newman2002spread}:
\begin{equation}
G_0(x;\widebar{T}) = \sum_{k=0}^\infty P_k \left(1-\widebar{T}+x \widebar{T}\right)^k\,.
\end{equation}
The distribution of occupied edges leaving a node at which we arrived by following a randomly selected edge is generated by~\cite{newman2002spread}
\begin{align}
G_1(x;\widebar{T})=\frac{G_0^\prime (x;\widebar{T})}{G_0^\prime (1;\widebar{T})} = \frac{\sum_{k=0}^\infty P_k k (1-\widebar{T}+x \widebar{T})^{k-1}}{\sum_{k=0}^\infty P_k k}\,.
\label{eq:generating_functions}
\end{align}
Note that we use $G_0'(x;\widebar{T})$ to indicate a derivative of $G_0(x;\widebar{T})$ with respect to $x$. To determine $\mathcal{S}(\widebar{T})$ (see Eq.~\eqref{eq:outbreak_size}), we solve the self-consistency equation 
\begin{equation}
u=G_1(u;\widebar{T})
\label{eq:self-consistency}
\end{equation}
and compute
\begin{equation}
\mathcal{S}(\widebar{T}) = 1-G_0(u;\widebar{T})\,,
\label{eq:outbreak_size}
\end{equation}
where $u$ is the probability that the node at the end of a randomly selected edge does not lead to a giant macroscopic component. In App.~\ref{app:correlated}, we generalize Eqs.~\eqref{eq:generating_functions} and \eqref{eq:outbreak_size} to account for correlation effects between nearest neighbors. Note that the generating function formalism is useful for cases when the exact network is unknown but only its degree distribution.

For details on limitations of the described bond-percolation mapping, see App.~\ref{sec:further_corrections} and Refs.~\cite{kenah2007second,trapman2007analytical,karrer2010message}.

As in Fig.~\ref{fig:hybrid_outbreak}, we now consider a random-regular graph with degree $k=5$. The degree distribution is $P_k = \delta_{k 5}$, where the Kronecker delta is $\delta_{k k'}=1$ if $k=k'$ and zero otherwise. In Fig.~\ref{fig:hybrid_outbreak}, we show the analytical solution of Eqs.~\eqref{eq:outbreak_size} and \eqref{eq:generating_functions} for fully Poissonian ($\widebar{T}_{\mathrm{PP}}$) and Poisson-geometric ($\widebar{T}_{\mathrm{PG}_1}$) SIR dynamics. For a small number of initially infected nodes, the relative outbreak size $S(\widebar{T})$ corresponds to the fraction of recovered nodes $\langle r \rangle$ (see Eq.~\eqref{eq:r_ensemble}). We observe that the bond-percolation description of hybrid and fully geometric SIR (see top left panel of Fig.~\ref{fig:hybrid_erlang}) outbreaks agree well with simulations.
\section{Unifying non-Markovian SIR processes}
\begin{figure}
\centering
\includegraphics[width=0.43\textwidth]{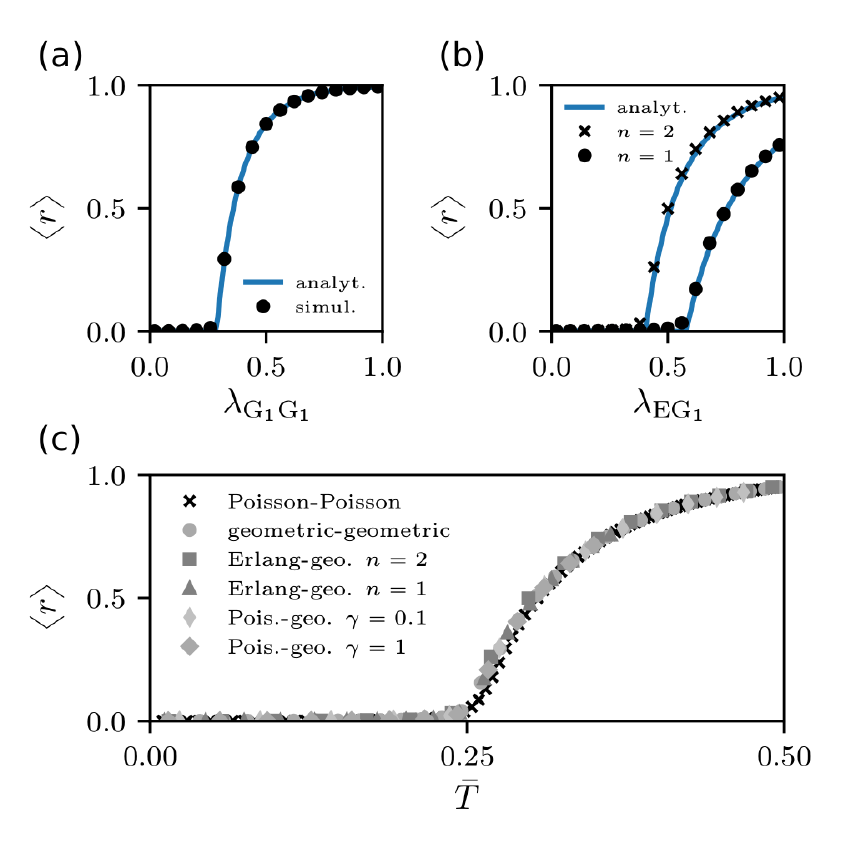}
\caption{\textbf{Outbreak sizes for non-Markovian SIR processes.} 
The fraction of recovered nodes $\langle r \rangle$ (see Eq.~\eqref{eq:r_ensemble}) for (a) geometric-geometric and (b) Erlang-geometric SIR processes as a function of the corresponding effective infection rates. (c) Markovian and non-Markovian outbreak sizes collapse onto the same curve when plotted against $\widebar{T}$.
Numerical simulations have been performed for $N=10^5$ nodes and $M=10^2$ realizations with 100 random initial infections. }
\label{fig:hybrid_erlang}
\end{figure}
Up to this point, we focused on hybrid SIR processes with variations in the infection-time distributions. To understand the general applicability of our framework, we now consider non-Markovian SIR dynamics with recovery times that are distributed according to the Erlang distribution
\begin{equation}
\phi(\tau)=\frac{\gamma^n \tau^{n-1} e^{-\gamma \tau}} {(n-1)!}\,,
\label{eq:gamma}
\end{equation}
where $n$ and $\gamma$ are the so-called shape and rate parameters. The Erlang distribution allows us to account for recovery processes that are not just exponentially distributed but more concentrated within a certain time window. It is the distribution that describes the sum of $n$ independent exponential variables with rate $\gamma$. The Erlang distribution has been used as an approximation of cell-cycle time distributions~\cite{Yates2017} and as such it is a good candidate for disease recovery processes as cells cycles have $n$ stages through which they are progressing (e.g., $n=4$ for COVID-19 \cite{Covid19}). In Fig.~\ref{fig:hybrid_erlang} (top right), we observe that Erlang-geometric SIR processes can also not be described by an effective infection rate. The considered examples of non-Markovian SIR processes show that the effective-spreading-rate description cannot uniquely characterize hybrid and general non-Markovian disease outbreaks. Instead, the mean transmissibility $\widebar{T}$ provides a unifying control parameter as we show in Fig.~\ref{fig:hybrid_erlang} (bottom). In App.~\ref{app:differentnetworks}, we outline that this also holds for other networks including Erd\H{o}s-R\'enyi, Barab\'asi-Albert, and various empirical networks.
\section{Discussion and conclusion}
We introduced numerical and analytical frameworks for the study of general (non-)Markovian SIR dynamics on networks. 
Furthermore, we proposed a novel hybrid SIR process that models infection and recovery as discrete-time Markovian and continuous-time (non-)Markovian processes, respectively. The discussed examples of hybrid SIR processes can account for cell-cycle distributions and latency intervals during which no infection events occur. We showed that the effective-infection-rate description of Markovian SIR processes~\cite{pastor2015epidemic} cannot uniquely capture non-Markovian epidemic outbreaks. However, our results suggest that the mean transmissibility provides a unifying description of (non-)Markovian SIR processes across a wide range of network structures (see App.~\ref{app:differentnetworks}) and infection and recovery time distributions. These observations are of particular interest for disease control and hint at a re-definition of the epidemic threshold to appropriately account for disease dynamics and network structure~\cite{liu2018measurability}. Our results also complement an earlier study on non-Markovian SIR dynamics~\cite{min2013suppression}, which showed that strong temporal heterogeneity in the contact patterns between individuals may significantly suppress epidemic outbreaks.

Further motivation for the study of discrete interaction processes comes from temporal-network theory, where the majority of temporal interactions is considered to be discrete~\cite{clauset2012persistence,holme2015modern}. Future studies may extend our work to hybrid models on temporal networks.

Our findings are in accordance with earlier results on non-Markovian susceptible-infected-susceptible (SIS) dynamics~\cite{starnini2017equivalence}, where a modified effective infection rate was used to uniquely capture corresponding steady states. A mean-field analysis of SIS dynamics~\cite{feng2019equivalence} also revealed that there is an equivalence between certain non-Markovian and Markovian SIS processes. Similar concepts may be helpful to better understand similarities between non-Markovian and Markovian SIR dynamics. 

To summarize, our work can contribute to more accurate and informative models of spreading processes on networks and meta-population spreading models~\cite{colizza2008epidemic,van2011gleamviz}.
\acknowledgments{We thank Jan Nagler for helpful comments. LB acknowledges financial support from the SNF Early Postdoc.Mobility fellowship on ``Multispecies interacting stochastic systems in biology'' and the Army Research Office (W911NF-18-1-0345). N.A-F. acknowledges financial support from ’SoBigData++’ with grant agreement 871042. L.B.~and N.A-F.~contributed equally to this work.
}
\clearpage
\appendix
\onecolumngrid
\section{Dynamics reconstruction}
\label{app:dynamics}
Based on the SPkMC framework that we outline in Sec.~\ref{sec:spkmc}, we can also determine the evolution of susceptible, infected, and recovered nodes. We use $1-\langle S(t) \rangle$ to denote the mean number of non-susceptible nodes prior to some time $t$ and compute this quantity from the set of edge-weighted spreading graphs $\lbrace G_1, \dots, G_M \rbrace$ according to
\begin{equation}
1 - \langle S(t) \rangle = \frac{1}{M} \sum_{i=1}^M \left| \bigcup_{k \in {\mathcal{I}}} \lbrace j: d_{{G}_i(k,j)} \leq t \rbrace \right|\,,
\label{eq:est1}
\end{equation}
%
%
where $\mathcal{I}$ is the set of initially infected nodes and $d_{G_i(k,j)}$ is the shortest-path length from node $k$ to node $j$ in the weighted spreading network $G_i$. To determine the fraction of infected and recovered nodes at time $t$, we compute the cardinality of the set of all nodes that are connected with an infected source node through a path of maximum length $t$. To extract the dynamical behavior of infected nodes from an SPkMC simulation, we need to include node-recovery weights $\lbrace \tau_i \rbrace$ in our simulation framework. Similarly to Eq.~\eqref{eq:est1}, we determine the mean number of infected nodes $\langle I (t) \rangle$ according to
\begin{equation}
\langle I(t) \rangle = \frac{1}{M} \sum_{i=1}^M \left| \bigcup_{k \in {\mathcal{I}}} \lbrace j: d_{G_{i(k,j)}} < t -\tau_j \rbrace \right|\,.
\end{equation}
\section{Quarantine dynamics}
\label{app:QuarantineDynamics}

\begin{figure}
\includegraphics[width=\textwidth]{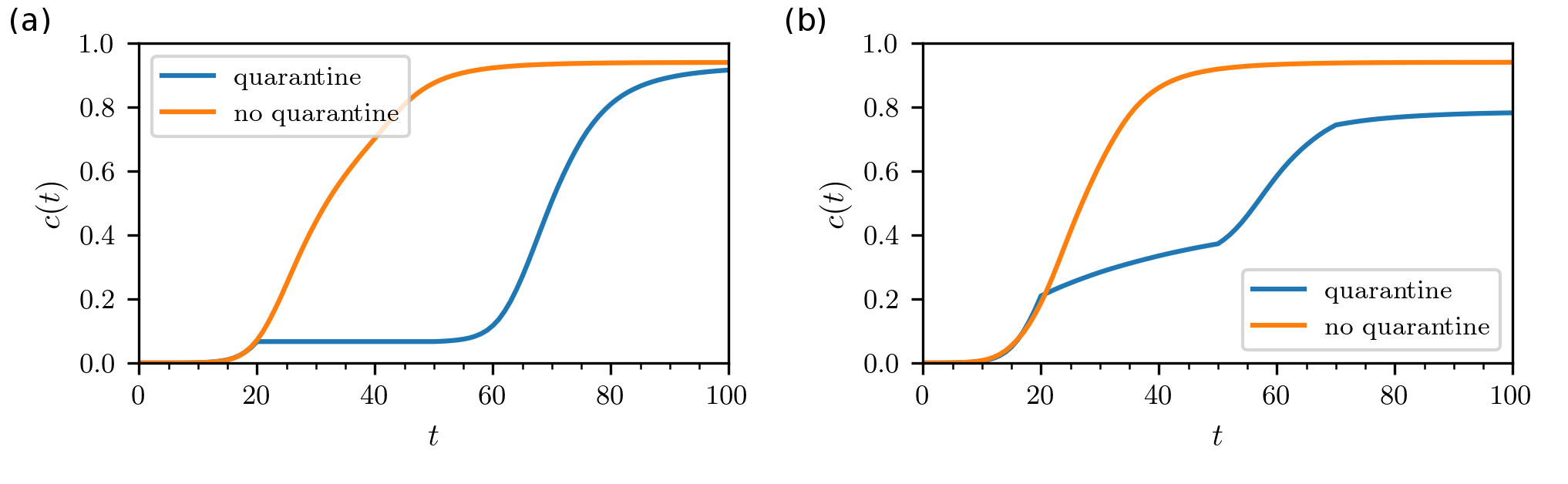}
\caption{\textbf{Quarantine with SPkMC simulations.} We use the SPkMC framework to simulate quarantine of nodes on a Erd\H{o}s-R\'enyi ($N=10^5$ nodes and mean degree $\langle k \rangle = 3$) for fully Poissonian SIR dynamics ($\beta=0.1, \gamma=0.02$). We show the cumulative proportion of infected nodes $c(t)=\int_0^t i(t')\,\mathrm{d}t'$ as a function of time $t$. (a) We quarantine all nodes from $t_1=20$ until $t_2=20+30=50$. (b) We quarantine 50\% of nodes from $t_1 = 20$ until $t_2 = t_1+30=50$ and from $t_3 = 70$ until $t_4 = t_3+30=100$.
}
\label{fig:quarantine}
\end{figure}

In this section, we describe the possibility to apply the SPkMC framework of Sec.~\ref{sec:spkmc} to quarantine modeling. Recall that Dijkstra's algorithm~\cite{dijkstra1959note} is constructing shortest paths via dynamic programming updates. At iteration $n=0$, all distances $d_{{G}_i(k,j)}[n]=\infty$ are set to infinity, except for the source node $d_{{G}_i(k,k)}[n]=0$. 
At iteration $n$, we update the shortest path with the following equation:
\begin{equation}
d_{{G}_i(k,l)}[n+1] = \min \left\{ d_{{G}_i(k,l)}[n], d_{{G}_i(k,j)}[n] + \rho_{j,l} \right\}\,,
\label{eq:quarantine}
\end{equation}
where $\rho_{j,l}$ is the edge weight between nodes $j$ and $l$.

One possible quarantine strategy would be that a pre-defined set of nodes gets removed from the network at quarantine time $t_1$ and brought back at time $t_2$. This procedure can be repeated as often as necessary and directly incorporated in our simulation framework. Nodes can only infect others or be infected by surrounding nodes if they are not under quarantine (i.e., not removed). The outlined quarantine protocol can be implemented as follows. For a given source node $k$, some target node $l$, and a node $j$ that is under quarantine during $t \in [t_1,t_2]$, the shortest-path calculation can be extended by adapting distance updates in Dijkstra's algorithm:

\begin{equation}
d_{{G}_i(k,l)}[n+1] = \min \left\{ d_{{G}_i(k,l)}[n], d_{{G}_i(k,j)}[n] + \rho_{j,l} + 
\chi_{B} ( d_{{G}_i(k,j)}[n] + \rho_{j,l})\right\}\,,
\label{eq:quarantine}
\end{equation}
where $\rho_{j,l}$ is the inter-event edge transmission delay between nodes, $B=\mathbb{R}_{\geq 0} \setminus [t_1,t_2]$, and $\chi_{B} (x)$ is the characteristic function of $B$:
\begin{equation}
\chi_{B} (x)= \begin{cases}
0\,,& x \in B\,,\\
+\infty\,,& x \notin B\,.
\end{cases}
\end{equation}
According to Eq.~\eqref{eq:quarantine}, we obtain a finite shortest-path length $d_{{G}_i(k,l)}$ if no quarantined node lies between nodes $k$ and $j$. That is, $d_{{G}_i(k,l)}$ is finite if $d_{{G}_i(k,j)} + \rho_{j,l} \in \mathbb{R}_{\geq 0} \setminus [t_1,t_2]$ for at least one node $j$. We show an SPkMC simulation for quarantine on a random network in Fig.~\ref{fig:quarantine}. We observe a drop in the number cumulative proportion of infections $c(t)=\int_0^t i(t')\,\mathrm{d}t'$ as soon as quarantine begins.

The prior quarantine protocol can be generalized as follows. Each node $j$ can have its own quarantine from time $t_1^{(j)}$ until $t_2^{(j)}$. Accordingly, we can define
$B^{(j)}=\mathbb{R}_{\geq 0} \setminus [t_1^{(j)},t_2^{(j)}]$ and $\chi_{B^{(j)}} (x)$, the characteristic function of $B^{(j)}$:
\begin{equation}
\chi_{B^{(j)}} (x)= \begin{cases}
0\,,& x \in B^{(j)}\,,\\
+\infty\,,& x \notin B^{(j)}\,.
\end{cases}
\end{equation}
Now, for a given source node $k$, some target node $l$, and a node $j$ that lies on the path between $k$ and $l$, the shortest-path calculation in Dijkstra's algorithm can be extended by adapting the dynamic programming update:
\begin{equation}
d_{{G}_i(k,l)}[n+1] = \min \left\{ d_{{G}_i(k,l)}[n], d_{{G}_i(k,j)}[n] + \rho_{j,l} + 
\chi_{B^{(j)}} ( d_{{G}_i(k,j)}[n] + \rho_{j,l}) + \chi_{B^{(l)}} ( d_{{G}_i(k,j)}[n] + \rho_{j,l})\right\}\,,
\label{eq:quarantine}
\end{equation}
where the characteristic function $\chi_{B^{(j)}}$ of node $j$ prohibits transmission \emph{from} node $j$ if it is under quarantine and the characteristic function $\chi_{B^{(l)}}$ prohibits transmission \emph{to} node $l$ if it is under quarantine.
\section{Comparison of algorithms}
\label{app:comparison}
\begin{figure}
\centering
\includegraphics[width=\textwidth]{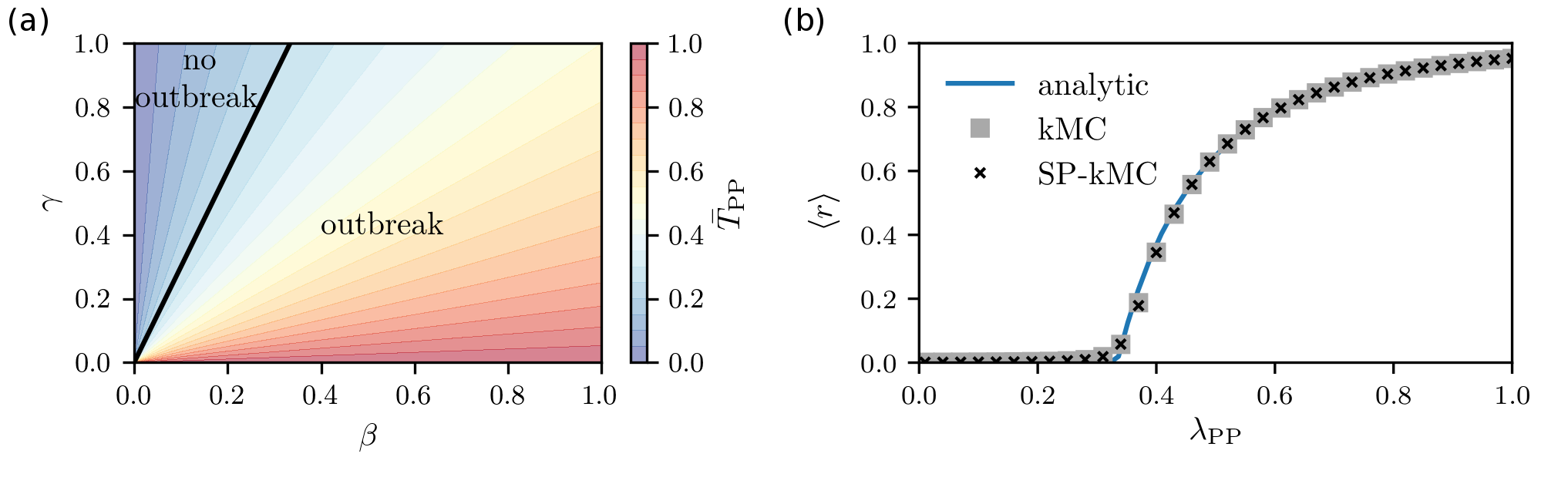}
\caption{\textbf{Stationary behavior in SPkMC and kMC SIR simulations.} (a) A contour plot of the Poissonian mean transmissibility (see Eq.~\eqref{eq:mean_Tapp}). The black solid line separates the phases with and without giant (outbreak) components (i.e., $\lambda_{\mathrm{PP}}^c=1/3$). (b) We consider a random-regular network with degree $k=5$ and $N=10^4$ nodes. For $\lambda_{\mathrm{PP}}>\lambda_{\mathrm{PP}}^c=1/3$ (see Eq.~\eqref{eq:lambda_capp}), the initial fraction of $10^{-3}$ infected nodes spreads through the system. We show the fraction of recovered nodes $\langle r \rangle$ as a function of $\lambda_{\mathrm{PP}}$. The blue solid line is a solution of SIR percolation problem~\cite{newman2002spread} (see Eqs.~\eqref{eq:self-consistency} and \eqref{eq:outbreak_size}) and the grey dots and black crosses are kMC and SPkMC simulations averaged over $10^3$ realizations (error bars are smaller than the markers).}
\label{fig:comparison}
\end{figure}
\begin{figure}
\includegraphics[width=\textwidth]{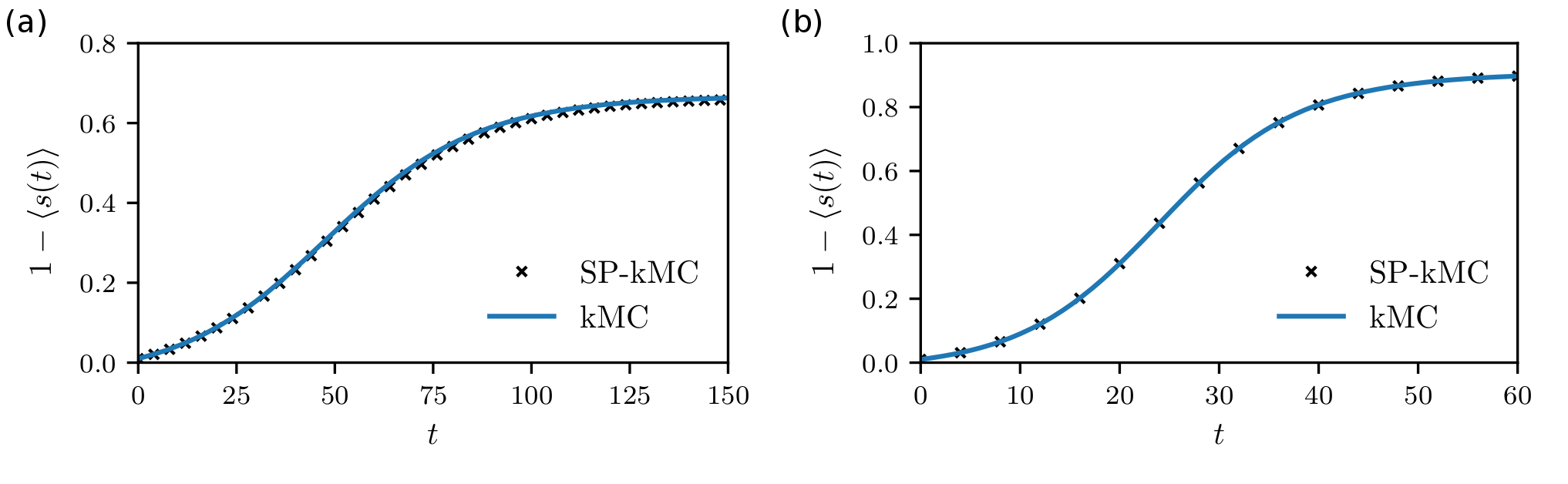}
\caption{\textbf{Transient behavior in shortest-path and kMC SIR simulations}. We show the fraction $1-\langle s(t) \rangle$ of non-susceptible nodes as a function of time for fully Poissonian SIR dynamics (blue solid line: kMC, black crosses: SPkMC). Simulations have been performed on a random-regular network with degree $k=5$ and $N=10^4$ nodes and the numerical data is based on $M=10^2$ samples. We used a recovery rate $\gamma=0.1$ and infection rates $\beta=0.05$ and $\beta=0.08$ in (a) and (b), respectively.}
\label{fig:compare_dynamics}
\end{figure}

Here, we compare the stationary and transient behavior of fully Poissonian SIR dynamics that we obtain with SPkMC and kMC simulations~\cite{istvan2019mathematics} (see Fig.~\ref{fig:comparison}). We consider a regular random network with degree $k=5$ and $N=10^4$ nodes. For Poissonian infection and recovery times, the mean transmissibility is~\cite{newman2002spread}
\begin{equation}
\widebar{T}_{\mathrm{PP}}=\widebar{T}_{\mathrm{PP}}(\lambda_{\mathrm{PP}}) = 1-\int_0^\infty \gamma e^{-\gamma \tau} e^{-\beta \tau} \, \mathrm{d}\tau=1-\frac{\gamma}{\beta+\gamma}=\frac{\lambda_{\mathrm{PP}}}{\lambda_{\mathrm{PP}}+1}\,,
\label{eq:mean_Tapp}
\end{equation}
where the effective infection rate is $\lambda_{\mathrm{PP}}=\beta/\mu$. This yields the threshold
\begin{equation}
\lambda_{\mathrm{PP}}^c=\frac{\langle k \rangle}{\langle k^2 \rangle - 2 \langle k \rangle}
\label{eq:lambda_capp}
\end{equation}
above which giant (outbreak) components are observable. We show a comparison of the transient behavior of SPkMC and kMC simulations in Fig.~\ref{fig:compare_dynamics}.
\section{Master equation for hybrid Poisson-geometric SIR dynamics}
\label{sec:master_equation}
In this section, we formulate the master equation for hybrid Poisson-geometric SIR dynamics. Let us denote the probability of finding a system in configuration $\sigma$ at time $t$ as $P(\sigma,t)$. The probabilities $P(\sigma,t)$ fulfill the normalization condition $\sum_{\sigma} P(\sigma,t)=1$ and the probabilistic evolution of the system is governed by the master equation:
\begin{equation}
\frac{\partial}{\partial t} P(\sigma,t)=\sum_{\sigma^{*} \neq \sigma } P(\sigma^{*}, t) W(\sigma^{*} \rightarrow \sigma) - \sum_{\sigma^{*}\neq \sigma }P(\sigma, t) W(\sigma \rightarrow \sigma^{*})\,.
\label{eq:master}
\end{equation}
The first term on the right-hand side of Eq.~\eqref{eq:master} describes the ``inflow'' into configuration $\sigma$ from other configurations $\sigma^*$ with transition rate $W(\sigma^{*} \rightarrow \sigma)$ and the second term accounts for the corresponding ``outflow'' with transition rate $W(\sigma \rightarrow \sigma^*)$. In the case of SIR dynamics, every configuration is a $n$-dimensional vector $\sigma=(\sigma_1, \dots, \sigma_n)$ that describes the state of every node with $\sigma_i \in \left\lbrace S,I,R \right\rbrace$. For hybrid Poisson-geometric SIR process, the transition rates are not constant in time anymore. This implies that the stochastic process is still Markovian, but not time-homogeneous as some transitions may only occur for integer times $t$. We factorize the transition rates in the following way:
\begin{equation}
W(\sigma^{*} \rightarrow \sigma,t) = \prod_{i=1}^n w\left( \sigma^{*}_i \rightarrow \sigma_i | \left\lbrace  \sigma^{*}_j \right\rbrace_{j:A_{i,j}=1}, t\right)\,,
\end{equation}
where $A$ is the adjacency matrix, $w(\cdot)$ denotes the local transition rate of state $\sigma_i$ conditioned on the neighboring states $\sigma_j^\ast$ at time $t$. The recovery rates are constant in time: 
\begin{equation}
w\left( \sigma^{*}_i = I \rightarrow \sigma_i = R | t \right) = \gamma\,.
\end{equation}
However, disease transmissions only occur when the time $t$ is an integer:
\begin{equation}
w\left( \sigma^{*}_i = S \rightarrow \sigma_i = I | \left\lbrace  \sigma^{*}_j = I \right\rbrace_{j:A_{i,j}=1}, t\right) \, \mathrm{d}t = \left( 1- \prod_{j:A_{i,j}=1} (1-q) \right) \mathbb{1}_{\mathbb{N}^+}(t)\,,
\end{equation}
where $\mathbb{1}_{\mathbb{N}^+}(t)$ denotes the indicator function of the positive natural numbers $\mathbb{N}^+$, which is equal to 1 when $t$ belongs to $\mathbb{N}^+$ and zero otherwise.
\section{Influence of discrete infection times on transmissibility}
\label{app:discrete_inf_times}
In Sec.~\ref{sec:hybrid}, we consider the following definition of the geometric distribution:
\begin{equation}
\psi_{\mathrm{G}_1}(\tau)=\sum_{k=1}^{\infty} \delta(\tau-k) (1-q)^{k-1}q\,,
\label{eq:geometric_app}
\end{equation}
where $q$ is the probability that an infection event occurs within a time interval of length 1 and $\delta(\cdot)$ is the Dirac delta function. This yields the mean transmissibility
\begin{align}
\begin{split}
\widebar{T}_{\mathrm{PG_1}} &= \widebar{T}_{\mathrm{PG_1}}(\gamma,q)= \int_1^\infty \phi(\tau) \sum_{\tau^\prime=1}^{\floor{\tau}} (1-q)^{\tau^\prime-1} q \,\mathrm{d}\tau\\
&=\int_1^\infty \phi(\tau) \left[1-(1-q)^{\floor{\tau}}\right]\,\mathrm{d}\tau \\
&= e^{-\gamma}-\int_1^\infty \phi(\tau) (1-q)^{\floor{\tau}}\,\mathrm{d}\tau\\
&= e^{-\gamma}-\sum_{k=1}^\infty \int_{k}^{k+1} \gamma e^{-\gamma \tau} (1-q)^{k} \, \mathrm{d} \tau\\
&= e^{-\gamma}-(e^{\gamma}-1)\sum_{k=1}^\infty (1-q)^k e^{-(k+1) \gamma} \\
&= e^{-\gamma}\left[1+\frac{\left(e^{\gamma }-1\right) (q-1)}{e^{\gamma }+q-1}\right]\,,
\end{split}
\end{align}
where $\phi(\tau)=\gamma e^{-\gamma \tau}$ is the exponential recovery time distribution with recovery rate $\gamma$. For small $\gamma$, the mean transmissibility is
\begin{equation}
\lim_{\gamma \rightarrow 0} \, \widebar{T}_{\mathrm{PG_1}}(\gamma, q)= \widebar{T}_{\mathrm{PG_1}}(\lambda_{\mathrm{PG_1}})= 1-\lambda_{\mathrm{P G_1}}^{-1}=\frac{\lambda_{\mathrm{P G_1}}-1}{\lambda_{\mathrm{P G_1}}}\,, 
\end{equation}
where $\lambda_{\mathrm{P G_1}}=1/(\gamma \langle \tau \rangle_\psi)=q/\gamma$. Based on the definition of the counting process in the Bernoulli trials, one can also define the geometric distribution where the counting starts at 1. There is no correct way, it depends on the actual definition of the stochastic process and alignment of the discrete and continuous inter-event times. Different definitions have different interpretations. For example, the above definition assumes that we allow the transmission to happen with probability $q$ at any non-negative integer number. However, if the counting process is defined over natural numbers (positive integers), it would represent the process with density
\begin{equation}
\psi_{\mathrm{G}_2}(\tau)= \sum_{k=0}^{\infty} \delta(\tau-k)(1-q)^{k} q
\label{eq:geometric2_app}
\end{equation}
and mean transmissibility
\begin{align}
\begin{split}
\widebar{T}_{\mathrm{PG_2}} &= \widebar{T}_{\mathrm{PG_2}}(\gamma, q)=\int_0^\infty \phi(\tau) \sum_{\tau^\prime=0}^{\floor{\tau}} (1-q)^{\tau^\prime} p \,\mathrm{d}\tau\\
&=\int_0^\infty \phi(\tau) \left[1-(1-q)^{1+\floor{\tau}}\right]\,\mathrm{d}\tau \\
&= 1-(1-q) \int_0^\infty \phi(\tau) (1-q)^{\floor{\tau}}\,\mathrm{d}\tau\\
&= 1- (1-q)\sum_{k=1}^\infty \int_{k-1}^k \gamma e^{-\gamma \tau} (1-q)^{k-1} \, \mathrm{d} \tau\\
&= 1-(e^{\gamma}-1)\sum_{k=1}^\infty (1-q)^k e^{-k \gamma} \\
&= 1+\frac{\left(e^{\gamma }-1\right)  (q-1)}{e^{\gamma }+q-1}\,.
\end{split}
\end{align}
\begin{figure}
\includegraphics[width=\textwidth]{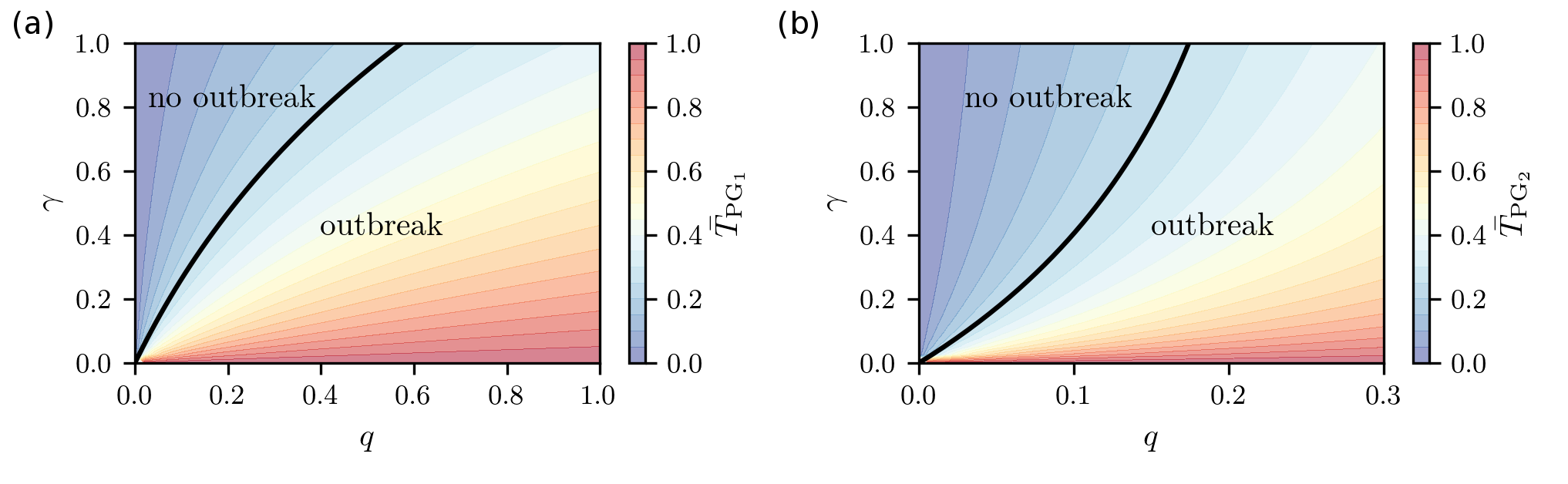}
\caption{\textbf{Influence of geometric distributions on transmissibility}. We illustrate the dependence of the mean transmissibility $\widebar{T}_{\mathrm{PG_1}}(\gamma,q)$ (a) and $\widebar{T}_{\mathrm{P G_2}}(\gamma,q)$ (b) on the infection probability $q$ and recovery rate $\gamma$. The black solid line separates the phases with and without giant (outbreak) components.}
\label{fig:phases}
\end{figure}
Note that the limiting behavior $\lim_{\gamma \to\infty} \widebar{T}_{\mathrm{PG_2}}(\gamma, q)= q$ is in sharp contrast to the Poissonian case, where $\lim_{\gamma \to\infty} \widebar{T}_{\mathrm{P P}}(\lambda_{\mathrm{PP}})= 0$. We find for small values of $q$ and $\gamma$:
\begin{equation}
\lim_{\gamma \rightarrow 0} \, \widebar{T}_{\mathrm{PG_2}}(\gamma, q)= \widebar{T}_{\mathrm{PG_2}}(\lambda_{\mathrm{PG_2}})= 1-\lambda_\mathrm{PG_2}^{-1}=\frac{\lambda_\mathrm{PG_2}-1}{\lambda_\mathrm{PG_2}}\,, 
\end{equation}
where $\lambda_\mathrm{PG_2}=1/(\gamma \langle \tau \rangle_{\psi})=q/\left[(1-q)\gamma\right]$. We show a comparison of the phase diagrams of hybrid SIR processes with infection-time distribution $\psi_{\mathrm{G}_1}$ and $\psi_{\mathrm{G}_2}$ in Fig.~\ref{fig:phases}.
\section{Correlated Networks}
\label{app:correlated}
The bond-percolation mapping that we outlined in Sec.~\ref{sec:bond_perc} is applicable to networks with an uncorrelated degree distribution (i.e., $P(k|k')=k P(k)/\langle k \rangle$). For networks with degree correlations, we use the notation~\cite{goltsev2008percolation} $P(k|k')=  \langle k \rangle P(k,k')/(k' P(k'))$ and extend the generating-function approach of Sec.~\ref{sec:bond_perc} according to
\begin{equation}
S(\widebar{T}) = 1-G_0(\mathbf{u};\widebar{T})\,,
\label{eq:outbreak_size_app}
\end{equation}
where $G_0(\mathbf{u};\widebar{T})=\sum_{k=0}^{k_\mathrm{cut}} P(k) \left(1-\widebar{T}+u_k \widebar{T}\right)^k$, $\mathbf{u}=(u_1,u_2,\dots,u_{k_\mathrm{cut}})$, $u_k=G_1(\mathbf{u};\widebar{T})$, and
\begin{align}
G_1(\mathbf{u};\widebar{T}) = \sum_{k'=0}^{k_\mathrm{cut}} \frac{\langle k \rangle P(k,k')}{k' P(k')} (1-\widebar{T}+u_{k'} \widebar{T})^{k'-1}\,.
\label{eq:generating_functions_app}
\end{align}
This extension accounts for correlations between neighboring nodes and is based on the assumption that the considered network is locally treelike~\cite{goltsev2008percolation}. We use $k_\mathrm{cut}$ to denote the largest degree in the network. The function $G_0(\mathbf{u};\widebar{T})$ is the generating function of the distribution of occupied edges belonging to a certain node. The distribution of occupied edges leaving a node at which we arrived by following a randomly selected edge is generated by $G_1(\mathbf{u};\widebar{T})$ and $u_k$ is the probability that a node with degree $k$ at the end of an randomly selected edge is occupied. Furthermore, note that degree-degree correlations become irrelevant for the percolation transition if the spectrum of the branching matrix $B_{k,k^{'}}=(k^{'}-1)P(k^{'}|k)$ satisfies certain conditions~\cite{goltsev2008percolation}. 

\begin{figure}
\centering
\includegraphics[width=0.87\textwidth]{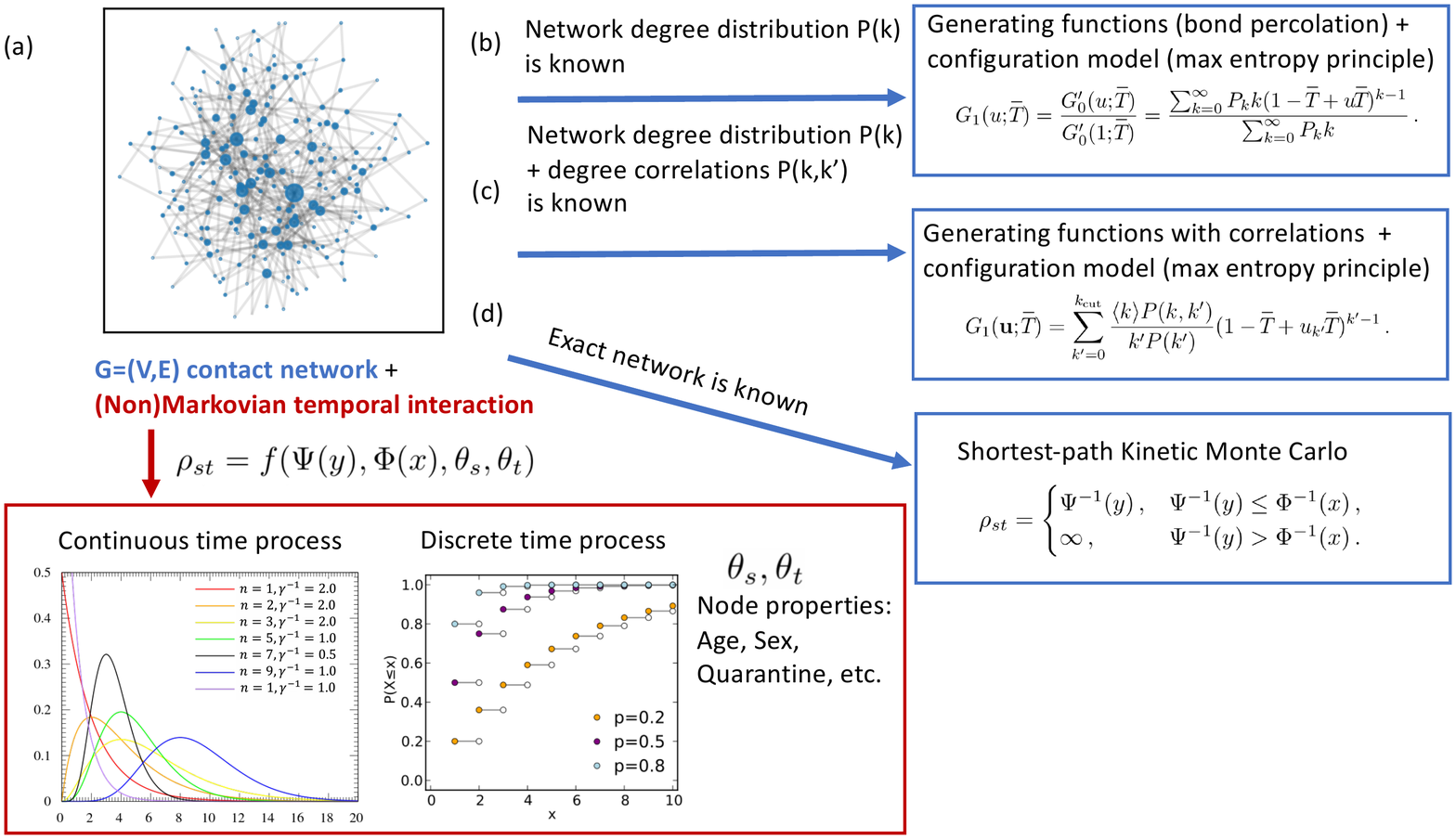}
\caption{Schematic diagram of proposed framework: (a) Example of a contact network (Barab\'asi-Albert) $N=225$ nodes and each new node is connected to $2$ existing nodes, node size scales with betweenness centrality. (b-c) Generating function formalism for a given network degree distribution with/without degree correlations. (d) SPkMC simulations on a given network. (e) (Non-)Markovian temporal interactions for generalized SIR process.}
\label{fig:schematic2}
\end{figure}

For a given network structure, we can use the SPkMC framework (see Sec.~\ref{sec:spkmc}) to simulate general (non-)Markovian SIR dynamics on a corresponding network. The generating-function approach provides a possibility to gain insights into (non-)Markovian disease outbreaks if we only know about the degree distribution of a certain network. In this case, the underlying network structure would be implicitly described by a configuration model, which corresponds to a network reconstruction using the max entropy principle~\cite{cimini2019statistical} with a certain degree distribution as constraint. We summarize these points in Fig.~\ref{fig:schematic2}.
\section{Different networks}
\label{app:differentnetworks}
\begin{figure}
\centering
\includegraphics[width=\textwidth]{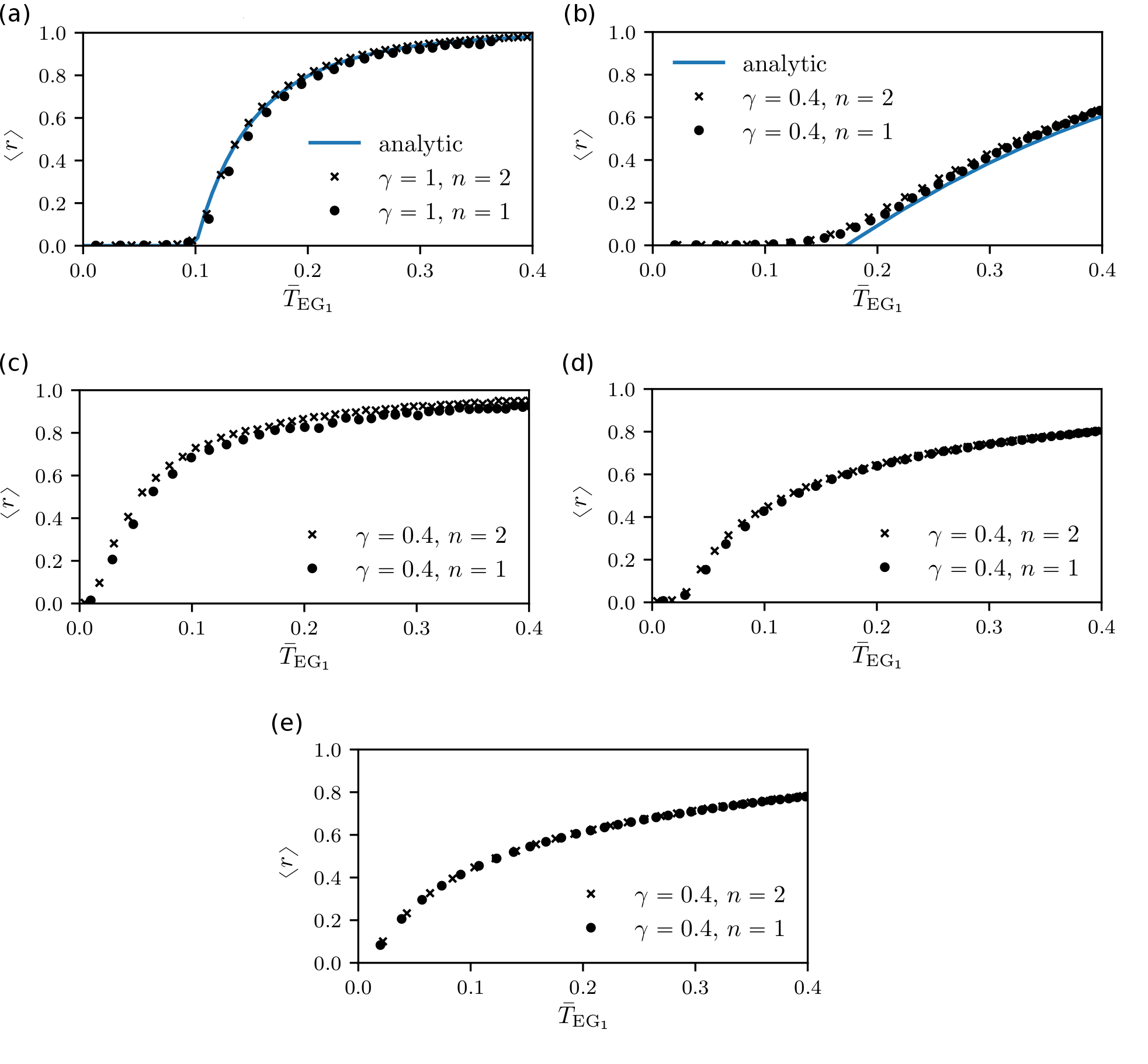}
\caption{\textbf{Phase diagrams of hybrid SIR processes on different networks}. We use the SPkMC framework to simulate Erlang-geometric SIR processes on different networks and show the corresponding phase diagrams (i.e., the fractions of recovered nodes $\langle r \rangle$ as a function of the mean transmissibility $\widebar{T}_{\mathrm{PG_1}}(\gamma,q)$. All simulations are based on 100 samples and a fraction of $0.01\%$ initially infected nodes. We used the following networks: (a) Erd\H{o}s-R\'enyi ($N=10^4$ nodes and mean degree $\langle k \rangle = 10$), (b) Barab\'asi-Albert ($N=10^4$ nodes and each new node is connected to $2$ existing nodes), (c) Facebook ($N=4039$ nodes and $\langle k \rangle=21.85$)~\cite{leskovec2012learning}, (d) Petster ($N=1858$ nodes and mean degree $\langle k \rangle=13.49$)~\cite{friendshipsapril}, and (e) LiveJournal ($N=5204176$ nodes and mean degree $\langle k \rangle = 18.90$)~\cite{mislove2007measurement}. The analytic solutions are based on Eqs.~\eqref{eq:self-consistency} and \eqref{eq:outbreak_size} (Erd\H{o}s-R\'enyi) and Eqs.~\eqref{eq:outbreak_size_app} and \eqref{eq:generating_functions_app} (Barab\'asi-Albert).}
\label{fig:phasediagrams}
\end{figure}
In the main text, we outlined that effective infection rates cannot uniquely capture hybrid and general non-Markovian disease outbreaks. However, our results for random-regular networks (see Fig.~\ref{fig:hybrid_erlang}) suggest that the mean transmissibility (see Eq.~\eqref{eq:transmissibility}) produces phase diagrams that are independent of underlying infection- and recovery-time distributions. In Fig.~\ref{fig:phasediagrams}, we show that this observation can also be made for other synthetic and real-world networks. For Erd\H{o}s-R\'enyi networks, we can use the bond-percolation description of disease outbreaks to analytically describe the phase diagram. In the case of Barab\'asi-Albert networks, we use the generating function approach for correlated networks (see App.~\ref{app:correlated}) to obtain the corresponding analytical description. 
\section{Further corrections to the percolation mapping}
\label{sec:further_corrections}
We utilized the SPkMC framework and Gillespie methods \cite{gillespie1976general,gillespie1977exact} to generate exact realizations of hybrid and non-Markovian SIR dynamics and compared them to corresponding analytical (bond percolation) predictions on synthetic and real-world networks. The discussed mapping to bond percolation can be further enhanced by introducing certain corrections, which is going to be part of our future work. For example, corrections to the mean transmissibility may result from considering the probability that $m$ out of $n$ edges get activated~\cite{tolic2018simulating}:
\begin{align}
\begin{split}
 T_{n,m} &= \binom{n}{m} 
 \int_0^\infty \phi(\tau) \left[ 1-\int_0^{\tau} \psi(t) \, \mathrm{d}t \right]^{n-m} \left[ \int_0^{\tau} \psi(t) \mathrm{d}t \right]^m \, \mathrm{d}\tau.
\end{split}
\end{align}
Note that the mean transmissibility of the main manuscript is a special case of $T_{n,m}$ for $m=n=1$. In addition, corrections accounting for semi-directed spreading involve tracking down the exact direction of activated links~\cite{kenah2007second} using an extended generating function formalism.
\end{document}